\documentclass[aps,pra,reprint,nofootinbib,showpacs,superscriptaddress]{revtex4-1}
\usepackage{graphicx}
\usepackage{enumerate}
\usepackage{hyperref}
\usepackage{amsfonts}
\usepackage{amsmath,amssymb}
\usepackage{mathrsfs}
\usepackage{amsthm}
\usepackage{dsfont}
\usepackage{bm}
\usepackage{color}
\usepackage{epstopdf}
\usepackage{epsfig}

{\rm }

\def\be{\begin{equation}}
 \def\ee{\end{equation}}
 \def\bea{\begin{eqnarray}}
 \def\eea{\end{eqnarray}}
 \def\bes{\begin{eqnarray}}
 \def\ees{\end{eqnarray}}
 \def\bi{\begin{itemize}}
 \def\ei{\end{itemize}} 

\renewcommand{\sec}[1]{\hyperref[sec:#1]{Sec.~\ref{sec:#1}}}

\newcommand{\fig}[1]{\hyperref[fig:#1]{Fig.~\ref{fig:#1}}}

\def\pa{\partial}

\def\2{\frac{1}{2}}
\def\4{\frac{1}{4}}

\begin{document}

\title{Continuous-time limit of topological quantum walks}

\author{Radhakrishnan Balu}
\affiliation{U.S.\ Army Research Laboratory, Adelphi, MD 20783, U.S.A.}
\author{Daniel Castillo}
\author{George Siopsis}
\affiliation{Department of Physics and Astronomy, The University of Tennessee, Knoxville, Tennessee 37996-1200, U.S.A.}
\author{Christian Weedbrook}
\affiliation{CipherQ, 10 Dundas St E, Toronto, M5B 2G9, Canada}

\date{\today}

\begin{abstract}

We derive the continuous-time limit of discrete quantum walks with topological phases.
We show the existence of a continuous-time limit that preserves their topological phases. We consider both simple-step and split-step walks, and derive analytically equations of motion governing their behavior. We obtain simple analytical solutions showing the existence of bound states at the boundary of two phases, and solve the equations of motion numerically in the bulk.

\end{abstract}

\maketitle

\section{Introduction}

A quantum walk~\cite{Kempe2003,SV,Kendon2006} is the quantum mechanical version of the classical walk and can either be discrete-time or continuous-time~\cite{Childs2010}.  Quantum walks  provide a fascinating and versatile framework for studying a myriad of physical processes ranging from biological systems~\cite{Mohseni2008} to satisfiability problems in computer science \cite {HEN} to universal quantum computation \cite{Child}.  A variety of experimental substrates have been used to implement proof-of-principle demonstrations of quantum walks; these include, single photons~\cite{Schreiber2010,Broome2010,Peruzzo2010,Schreiber2012,Sansoni2012,Jeong2013}, trapped ions~\cite{Zahringer2010,Schmitz2009}, and atoms in optical lattices~\cite{Karski2009,Genske2013}. 

One of the most fascinating aspects of quantum walks is related to the studying of topological phases in solid state physics~\cite{Hasan2010,Qi2011}. It was shown~\cite{Kit,Kitagawa2012,Kitagawa2012b} that the discrete version of quantum walks can be used to model and understand topological phases. This initial work has been theoretically extended~\cite{Lindner2011,Obuse2011,Asboth2012,Asboth2013,Ramasesh2016,Rakovszky2015} and also experimentally demonstrated~\cite{Cardano2016}.

In this work, we consider the continuous-time limit of discrete topological quantum walks. This is investigated with simple and split-step evolutions in which we identify their phases and boundaries. We confine ourselves to space-discrete and time-continuous walks which become relativistic in space limits~\cite{Strauch}. One way to create non-trivial topological phases is to fashion the quantum walks in split steps using two coins with a requirement that the second coin's rotation is inhomogeneous~\cite{Kit}. 

We show that even with simple evolution one can have topologically bound states by considering an equivalent representation that splits the single coin's rotation in two steps. This is then generalized to split step with an additional coin providing a rich structure to the quantum walks adding new phases and boundaries. A key consideration in formulating continuous-time quantum walks from their discrete counterparts is to take the limit in the product topology of the spaces $\theta\in\mathscr{R}\times{R}\ni{\omega}$, where, $\theta$ represents the rotation angle and $\omega$ enters the limiting process as we study the systems in momentum space.

Our discussion is organized as follows. In Section \ref{sec:1}, we analyze the simple-step walk initially from a discrete-limit walk, and then from a continuous-time limit, perspective. In Section \ref{sec:2}, we consider the case of split-step walk again with both a discrete-time walk and a continuous-time walk. We offer conclusions in Section \ref{sec:3}.

\section{Simple-step Walk}
\label{sec:1}

Before we introduce the continuous-time limit for the simple-step walk, we rederive the discrete case. These dicsrete results were originally presented in~\cite{Kit,Kitagawa2010,Obuse2015}. We do this for completeness for the reader and also to show that our results match, so we have a good limit in the continuum.

\subsection{Discrete-time walk}
Consider a system consisting of a quantum coin and a quantum walker. The quantum coin is a qubit in $\mathcal{H}_c$, whereas the Hilbert space of the quantum walker $\mathcal{H}_w$ is infinite-dimensional. Let $\{ |0\rangle_c , |1\rangle_c \}$ and $\{ |x\rangle_w, x\in\mathbb{Z}\}$ be orthonormal bases in $\mathcal{H}_c$ and $\mathcal{H}_w$, respectively. The integer $x$ can be thought of as the position of the (classical) walker. A general state can be written as
\be |\Psi\rangle = \sum_x \left[ \Psi_0(x) |0\rangle_c + \Psi_1(x) |1\rangle_c \right] |x\rangle_w
\ee
In momentum space, we define
\be |k\rangle_w = \frac{1}{\sqrt{2\pi}} \sum_x e^{-ikx} |x\rangle_w \ , \ \ k \in (-\pi, \pi) \ee
We deduce
\be \langle k |k'\rangle = \delta (k-k') \ , \ \ |x\rangle = \frac{1}{\sqrt{2\pi}} \int_{-\pi}^\pi dk\ e^{ikx} |k\rangle \ee
Introduce operators which shift the position of the walker,
\be L^\pm |x\rangle_w = |x\pm 1\rangle_w \ee
and a unitary representing the flipping of the coin,
\be\label{eqTtheta} T(\theta) = \begin{pmatrix}
	\cos\theta & \sin\theta \\ \sin\theta & - \cos\theta
\end{pmatrix}
= e^{-i\theta Y} Z~.
\ee
Subsequently to coin flipping, the walker moves forward if the coin is in $|0\rangle$, and backwards if the coin is in $|1\rangle$. This is represented by the operator
\be S = \Pi_0 \otimes L^+ + \Pi_1 \otimes L^- \ , \ \ \Pi_i = |i\rangle\langle i| \ \ \ (i=0,1) \ee
Thus a step of the walker is represented by
\be U = ST(\theta) = S e^{-i\theta Y}  Z~. \ee
After $n$ steps, the initial state evolves with $U^n$.
In this chain, it is convenient to define the repeated block
\be\label{eq10a} U' = e^{ -i\frac{\theta}{2} Y}  Z  S e^{-i \frac{\theta}{2} Y} \ee
The advantage of working with $U'$ instead of $U$ is that $U'$ is of the form
\be\label{eq10} U' = iF X F^{-1} X \ee
where
\be F = e^{-i \frac{\theta}{2} Y} e^{-i\frac{\pi}{4} Z} S_-
\ee
and
\be S_+ = \Pi_0 \otimes L^+ + \Pi_1 \otimes \mathbb{I} \ ,\ \
S_- = \Pi_0 \otimes \mathbb{I} + \Pi_1 \otimes L^- \ee
$X$
acts as a ``chiral" transformation ($X e^{-i\theta Y} X = e^{i\theta Y}$, and $X Z X = -Z$).

To show \eqref{eq10}, notice that $Z$ commutes with $S_\pm$, and
\be F^{-1} =  X S_+ X e^{i\frac{\pi}{4} Z}
e^{i\frac{\theta}{2} Y}~, \
S = S_- S_+~. \ee
To find the eigenvalues and corresponding eigenstates, it is convenient to switch to the frame in which $X$ is diagonal. This can be accomplished with $e^{i\frac{\pi}{4} Y}$, since we have
$e^{i\frac{\pi}{4} Y} X e^{-i\frac{\pi}{4} Y} = Z$. We will therefore work with
\be\label{eq13} W = e^{i\frac{\pi}{4} Y}  U'  e^{-i\frac{\pi}{4} Y} \ , \ \ G = e^{i\frac{\pi}{4} Y}  F  e^{-i\frac{\pi}{4} Y} \ee
To solve the eigenvalue problem, notice that
the momentum eigenstates are eigenstates of the shift operators,
\be L^\pm |k\rangle_w = e^{\pm ik} |k\rangle_w \ee
Therefore,
\be\label{eq15} W |k\rangle_w = W_c
|k\rangle_w \ , \ \ G |k\rangle_w = G_c
|k\rangle_w\ee
where the matrices $W_c$ and $G_c$ act on the coin. We obtain
\be W_c
= \begin{pmatrix}
	i\sin k \cos\theta & -\cos k -i \sin k \sin\theta \\ -\cos k +i \sin k \sin\theta & i \sin k \cos\theta
\end{pmatrix} \ee
and
\be G_c
= e^{i\frac{\pi}{4} Y} \begin{pmatrix}
	e^{-i\frac{\pi}{4} } \cos\frac{\theta}{2} & -e^{i\frac{\pi}{4} } e^{-ik} \sin\frac{\theta}{2} \\ e^{-i\frac{\pi}{4} } \sin\frac{\theta}{2} & e^{i\frac{\pi}{4} } e^{-ik} \cos\frac{\theta}{2}
\end{pmatrix} e^{-i\frac{\pi}{4} Y} \ee
%
%
The eigenvalues of $W_c$ are
\be e^{-i\omega_\pm (k)} =  i\cos\theta \sin k \mp  \sqrt{1 - \cos^2\theta \sin^2 k} \ee
Note that $\sin\omega_\pm = \mp\cos\theta \sin k$.

We introduce two topological invariants,
\be \nu_\alpha = \frac{1}{2\pi i} \int_{-\pi}^{\pi} dk \frac{d}{dk} \ln {}_c\langle\alpha|G_c|0\rangle_c \ , \ \ \alpha = 0,1 \ee
Explicitly,
\be \nu_0 =  \int_{-\pi}^{\pi} \frac{dk}{2\pi} \frac{1}{1 - i z e^{ik}} \ , \ \
z = \frac{1+\tan \frac{\theta}{2}}{1-\tan \frac{\theta}{2}} \ee
We obtain
\be \nu_0 = \left\{ \begin{array}{ccc}
	1 & , & \theta > 0 \\ 0 & , & \theta < 0
\end{array}\right. \ee
Similarly,
\be \nu_1 =  \int_{-\pi}^{\pi} \frac{dk}{2\pi} \frac{1}{1 + i e^{ik} /z} = \left\{ \begin{array}{ccc}
	0 & , & \theta > 0 \\ 1 & , & \theta < 0
\end{array}\right. \ee
Thus, we have two distinct topological phases with $\nu_0 = 1$, $\nu_1 = 0$ ($\nu_0 = 0$, $\nu_1 = 1$) for $\theta > 0$ ($\theta <0$).
%

In position space, the action of $W$ can be deduced directly from \eqref{eq10a} and \eqref{eq13}. We obtain
\bea\label{eq8} \Psi_0 (x) &\to& \frac{\cos\theta}{2}    \left[ \Psi_0 (x-1) - \Psi_0 (x+1) \right] \nonumber\\
&& - \frac{1 + \sin\theta}{2}  \Psi_1 (x-1) - \frac{1 -  \sin\theta}{2}  \Psi_1 (x+1)\nonumber\\
\Psi_1 (x) &\to& \frac{\cos\theta}{2}   \left[ \Psi_1 (x-1) - \Psi_1 (x+1) \right] \nonumber\\
&& - \frac{1 - \sin\theta}{2}  \Psi_0 (x-1) - \frac{1 + \sin\theta}{2}  \Psi_0 (x+1)
\nonumber\\
\eea

\begin{figure}[htp]
	\centering
	\includegraphics[scale=1.5]{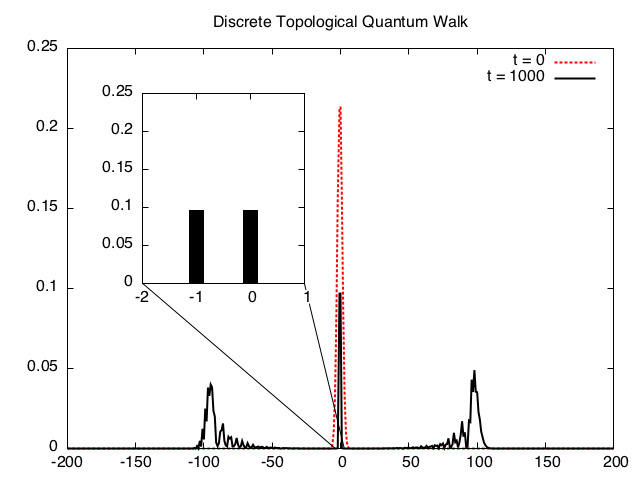}
	\caption{Trapped state at the boundary of two topological phases of a discrete simple-step walk.}\label{fig:1}
\end{figure}

\begin{figure}[htp]
	\centering
	\includegraphics[scale=0.4]{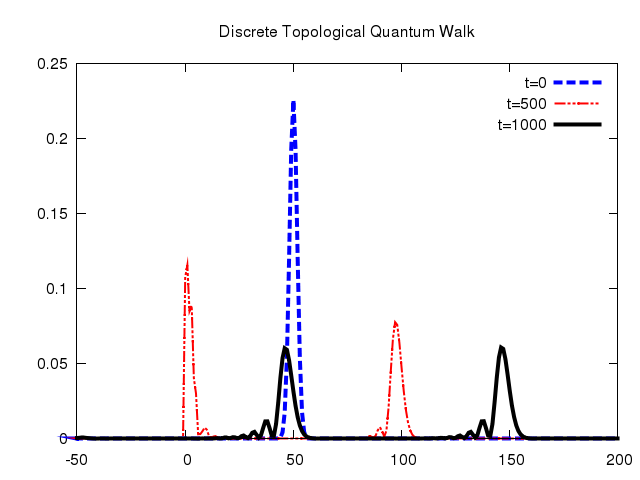}
	\caption{State being reflected at the boundary of two topological phases of a discrete simple-step walk.}\label{fig:2}
\end{figure}

In Fig.~\ref{fig:1}, we see that a discrete-time simple-step topological walk gives rise to bound states at $x=0$ and $x=-1$, which is chosen to be the boundary of two distinct topological phases. The initial state was centered around $x=0$ with a small spread $\Delta x$, and $\Psi_0(x) = \Psi_1(x)$. As the system evolves in time, the parts of the state near the boundary diffuse ballistically away from the boundary, but the part at the boundary remains protected. In Fig.~\ref{fig:2}, we choose the initial state around $x=50$, i.e., entirely within a single topological phase. In this case, we see that the quantum walk diffuses in both directions ballistically. The part of the walk that reaches the boundary of the other phase is reflected back and continues to diffuse away from the boundary ballistically without entering the region of the other topological phase.

\subsection{Continuous-time limit}

To go over to the continuous-time limit, we set $\theta = \frac{\pi}{2} - \epsilon$, and consider a scaling limit in which $\epsilon \to 0$ and $n\to\infty$, so that the product $n\epsilon $ remains finite. In this limit, $\omega_+ \to \pi$, and $\omega_- \to 0$.
Notice that $W^2 = \mathbb{I} + \mathcal{O} (\epsilon)$, which is not the case for $W$. We will therefore consider the limit of an \emph{even} number of steps.

Setting
\be\epsilon = \gamma\Delta t \ , \ \ t = 2n\Delta t \ee
Applying \eqref{eq8} twice, we obtain
\bea\label{eq8a} &&\Psi_0 (x, n+2) - \Psi_0 (x,n) \nonumber\\
&& = \gamma \Delta t \left[  \Psi_1 (x,n) - \Psi_1 (x-2,n) \right] \nonumber\\
&& \Psi_1 (x, n+2) - \Psi_1 (x,n) \nonumber\\
&&=  -\gamma \Delta t \left[  \Psi_0 (x,n) - \Psi_0 (x+2,n)  \right]
\eea
from which we deduce the continuous-time quantum walk in the limit $\Delta t\to 0$,
\bea\label{eq8b}  \frac{\partial\Psi_0 (x, t)}{\partial t} &=& \gamma  \left[
\Psi_1 (x,t) -  \Psi_1 (x-2,t) \right] \nonumber\\
\frac{\partial\Psi_1 (x, t)}{\partial t} &=& \gamma  \left[
- \Psi_0 (x,t) +  \Psi_0 (x+2,t) \right]
\eea
Defining
\be\label{eq27} \Phi_\pm (x) = \pm \Psi_0 (x) +  \Psi_1 (x-1 ) \ee
we obtain the decoupled equations
\be\label{eq37} \frac{\partial\Phi_\pm (x, t)}{\partial t} = \pm \gamma  \left[
\Phi_\pm (x+1,t) - \Phi_\pm (x-1,t) \right]
\ee
related to each other by time reversal.

Working similarly in the other phase (with $\nu_0 = 0$), we obtain in the continuous-time limit,
\bea\label{eq8bb}  \frac{\partial\Psi_0 (x, t)}{\partial t} &=& \gamma  \left[
\Psi_1 (x,t) -  \Psi_1 (x+2,t) \right] \nonumber\\
\frac{\partial\Psi_1 (x, t)}{\partial t} &=& \gamma  \left[
- \Psi_0 (x,t) +  \Psi_0 (x-2,t) \right]
\eea
It is easy to see that these are equivalent to the decoupled Eqs.\ \eqref{eq37} under the definition $\Phi_\pm (x) = \mp \Psi_0 (x) +  \Psi_1 (x+1 )$ (\emph{cf.}\ with Eq.\ \eqref{eq27}).

The above results are no longer valid if the coin parameters are spatially dependent. In particular, we are interested in the case in which there are two regions in space which have different topological numbers. This can be achieved by having $\theta > 0$ and $\theta < 0$ in the respective regions.
Then the above results are valid in the bulk of each region, but not along their boundaries.

For the continuous-time limit, we need to consider the limit of $4n$ steps as $n\to\infty$. This is because $W^2 \ne \mathbb{I} + \mathcal{O} (\epsilon)$, due to an obstruction at the boundary of the two regions, but we still have $W^4 = \mathbb{I} + \mathcal{O} (\epsilon)$.


\begin{figure}[htp]
	\centering
	\includegraphics[scale=1.5]{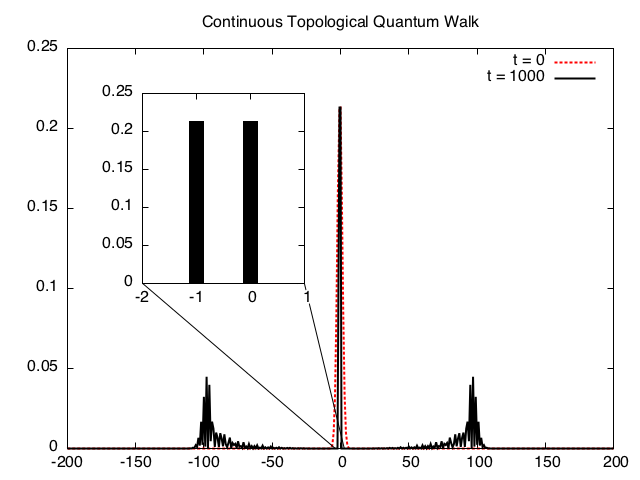}
	\caption{Trapped state at the boundary of two topological phases of a continuous simple-step walk.}\label{fig:3}
\end{figure}

\begin{figure}[htp]
	\centering
	\includegraphics[scale=0.4]{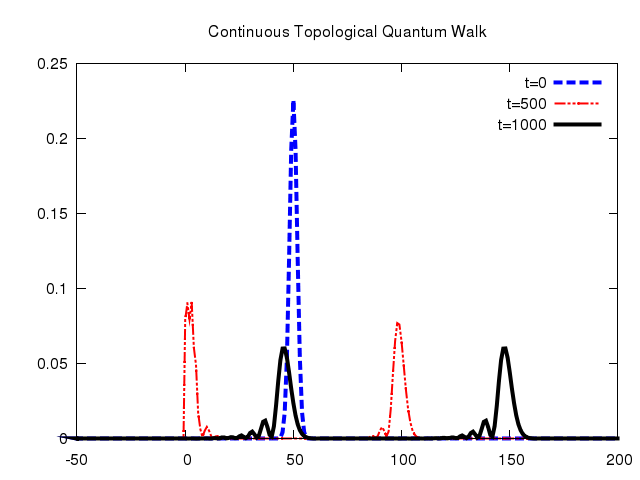}
	\caption{State being reflected at the boundary of two topological phases of a continuous simple-step walk.}\label{fig:4}
\end{figure}
In the continuous-time limit, away from the boundary the results match those of single topological phases. More precisely, for $x\ge 2$, we recover Eq.\ \eqref{eq8bb} whereas for $x\le -3$, we recover Eq.\ \eqref{eq8b}.
Near the boundary (for $-3 < x < 2$), we obtain
\bea \frac{\partial\Psi_0(1)}{\partial t}
&=& \gamma  \Psi_1 (1) \nonumber
\\ \frac{\partial\Psi_1 (1)}{\partial t}  &=& - \gamma \left[  \Psi_0 (1) -  \Psi_0 (3) \right]  \nonumber
\\ \frac{\partial\Psi_0(0)}{\partial t}
&=& 0 \nonumber\\
\frac{\partial\Psi_1 (0)}{\partial t}  &=&  \gamma  \Psi_0 (2)  \nonumber
\\
\frac{\partial\Psi_0 (-1)}{\partial t}  &=&  0
\\
\frac{\partial\Psi_1(-1)}{\partial t}
&=& \gamma  \Psi_0 (-3)  \nonumber \nonumber\\
\frac{\partial\Psi_0 (-2)}{\partial t}  &=& \gamma  \Psi_1 (-2)  \nonumber\\
\frac{\partial\Psi_1(-2)}{\partial t}
&=& -\gamma \left[  \Psi_0 (-2) -  \Psi_0 (-4) \right]
\eea
Notice that $\Psi_0 (0)$ and $\Psi_1 (-1)$ decouple, so if initially $\Psi_0(0) =1$, then the walk is trapped at $x=0$, and similarly for $\Psi_1(-1)$. Moreover, if a walk starts entirely within one of the topologically distinct regions (or at the boundary, $x=0$), it remains in it. This matches the asymptotic behavior observed in the discrete-time case (cf. Fig.~\ref{fig:1}).

By defining
$\Phi_\pm$ as in \eqref{eq27}, we recover the equations of motion for decoupled walks
\eqref{eq37} away from the boundary (for $|x|\ge 2$). In the continuous-time limit of the simple-step quantum walk, we see the same behavior as we did for the discrete quantum walk. We have the same two bound states near the boundary at $x = 0,1$, respectively. This is shown in Fig.~\ref{fig:3}. Away from the boundary, the system diffuses ballistically, and is reflected at the boundary, as shown in Fig.~\ref{fig:4}.

\section{Split-step Walk}
\label{sec:2}

As with the previous section we rederive the discrete-time case here for completeness. The original results can be found in \cite{Kit,Kitagawa2010,Obuse2015} along with our new results in the continuous-time limit following thereafter.

\subsection{Discrete-time walk}
For the split-step walk, we flip two coins, $T(\theta_1)$ and $T(\theta_2)$, where
$T(\theta)$ is defined in \eqref{eqTtheta}.
A step of the walker is represented by
\be U = S_-T(\theta_2)S_+ T(\theta_1) \ee
It is convenient to define the repeated block
\be\label{eq10as} U' = e^{-i \frac{\theta_1}{2} Y} Z  S_- e^{-i \theta_2 Y} Z S_+ e^{-i\frac{\theta_1}{2} Y} \ee
The advantage of working with $U'$ instead of $U$ is that $U'$ is of the form
\be\label{eq10s} U' = -F X F^{-1} X \ee
where
\be\label{eqFsplit} F = e^{-i\frac{\theta_1}{2} Y} ZS_- e^{-i\frac{\theta_2}{2} Y}
\ee
After switching to a frame in which $X$ is diagonal, we arrive at $W$ given by \eqref{eq13}, and $W_c$ acting on a coin (Eq.\ \eqref{eq15}) as
\be W_c
= \begin{pmatrix}
	\beta_0 & \beta_1 \\ -\beta_1^\ast & \beta_0
\end{pmatrix} \ee
where
\bea \beta_0 &=& \cos k \cos\theta_1\cos\theta_2 + \sin\theta_1\sin\theta_2 \nonumber\\
\beta_1 &=& -(i\sin k + \cos k \sin\theta_1)\cos\theta_2 + \cos\theta_1\sin\theta_2\ \ \
\eea
The eigenvalues are
\be e^{-i\omega_\pm} = \beta_0 \mp i\sqrt{1-\beta_0^2}  \ee
Similarly, for $G_c$ defined by \eqref{eq13} and \eqref{eq15} with $F$ given by
\eqref{eqFsplit}, we obtain
\be G_c
= e^{-ik/2}\begin{pmatrix}
	\gamma_0 & \gamma_1 \\ \gamma_1^\ast & -\gamma_0^\ast
\end{pmatrix} \ee
where
\bea \gamma_0 &=& \cos \frac{k}{2} \cos\frac{\theta_-}{2} + i \sin \frac{k}{2} \cos\frac{\theta_+}{2} \nonumber\\
\gamma_1 &=& \cos \frac{k}{2} \sin\frac{\theta_-}{2} - i \sin \frac{k}{2} \cos\frac{\theta_+}{2}\ \ \
\eea
and we defined $\theta_\pm = \theta_1 \pm \theta_2$.

For the two topological invariants, we obtain, respectively,
\be \nu_0 = \int_{-\pi}^{\pi} \frac{dk}{2\pi}\frac{1}{1-z_0 e^{ik} } \ , \ \ z_0 = \frac{\cos\frac{\theta_+}{2} + \sin\frac{\theta_-}{2}}{\cos\frac{\theta_+}{2} - \sin\frac{\theta_-}{2}} \ee
and
\be \nu_1 = \int_{-\pi}^{\pi} \frac{dk}{2\pi}\frac{1}{1+ z_1 e^{ik} } \ , \ \ z_1 = \frac{\cos\frac{\theta_-}{2} - \sin\frac{\theta_+}{2}}{\cos\frac{\theta_-}{2} + \sin\frac{\theta_+}{2}}
\ee
It is easy to see that
\be \nu_\alpha = \left\{ \begin{array}{ccc}
	1 & , & |z_\alpha| < 1 \\ 0 & , & |z_\alpha|>1
\end{array}\right. \ , \ \
\alpha = 0,1~.
\ee
Thus we obtain four different topological phases,
\begin{align}
	I (\nu_0,\nu_1) &= (1, 1)\\
	II (\nu_0,\nu_1) &= (0, 0)\\
	III (\nu_0,\nu_1)  &= (0,1)\\
	IV (\nu_0,\nu_1) &= (1,0)
\end{align}
In position space, the action of $W$ yields
\begin{widetext}
\bea\label{eq8split} \Psi_0 (x) &\to& \frac{\cos\theta_1\cos\theta_2}{2}    \left[ \Psi_0 (x-1) + \Psi_0 (x+1) \right] + \sin\theta_1\sin\theta_2 \Psi_0 (x)\nonumber\\
&& - \frac{1 + \sin\theta_1}{2}\cos\theta_2  \Psi_1 (x-1) + \frac{1 -  \sin\theta_1}{2}\cos\theta_2  \Psi_1 (x+1) + \cos\theta_1 \sin\theta_2 \Psi_1 (x)\nonumber\\
\Psi_1 (x) &\to& \frac{\cos\theta_1\cos\theta_2}{2}    \left[ \Psi_1 (x-1) + \Psi_1 (x+1) \right] + \sin\theta_1\sin\theta_2 \Psi_1 (x) \nonumber\\
&& - \frac{1 - \sin\theta_1}{2}\cos\theta_2  \Psi_0 (x-1) + \frac{1 +  \sin\theta_1}{2}\cos\theta_2  \Psi_0 (x+1) - \cos\theta_1 \sin\theta_2 \Psi_0 (x)
\eea
\end{widetext}

\begin{figure}[htp]
	\centering
	\includegraphics[scale=1.5]{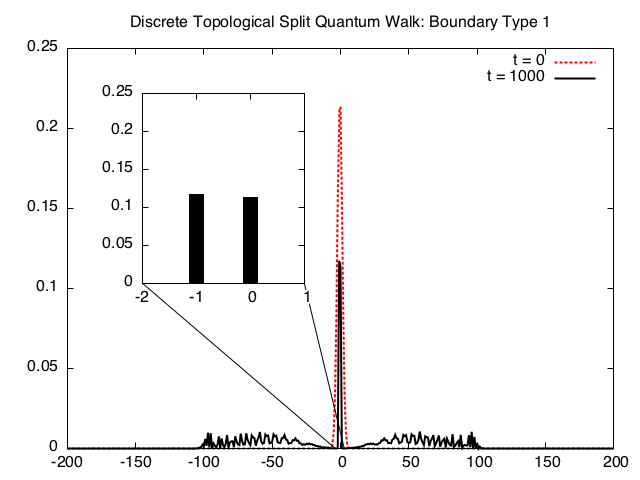}
	\caption{Trapped state at the boundary of topological phases \emph{I} and \emph{II} of a continuous simple-step walk.}\label{fig:5}
\end{figure}

\begin{figure}[htp]
	\centering
	\includegraphics[scale=0.35]{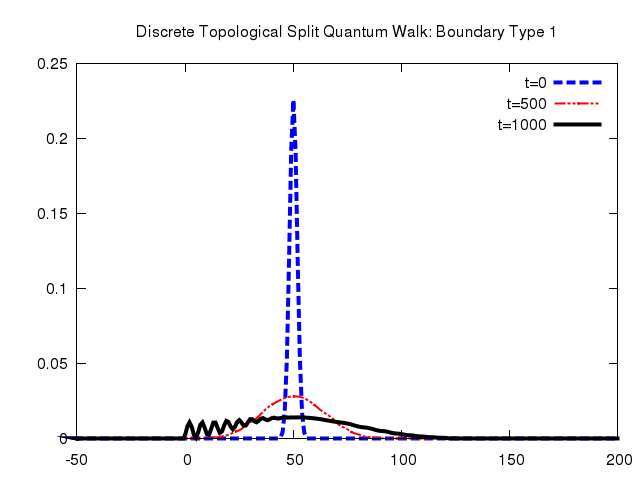}
	\caption{State being reflected at the boundary of topological phases \emph{I} and \emph{II} of a continuous simple-step walk.}\label{fig:7}
\end{figure}

\begin{figure}[htp]
	\centering
	\includegraphics[scale=1.5]{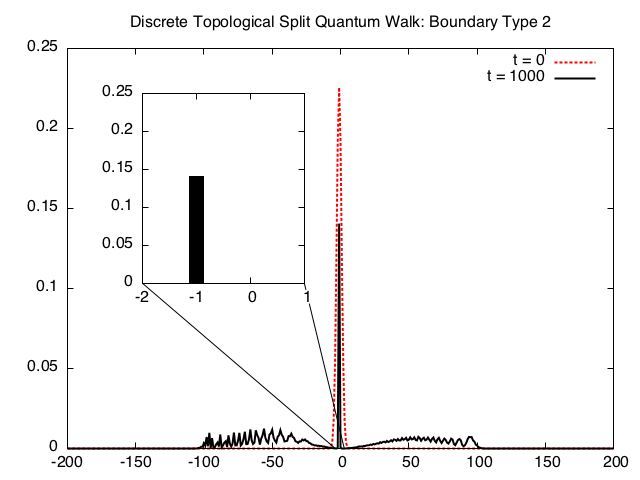}
	\caption{Trapped state at the boundary of topological phases \emph{I} and \emph{III} of a continuous simple-step walk.}\label{fig:6}
\end{figure}

\begin{figure}[htp]
	\centering
	\includegraphics[scale=0.35]{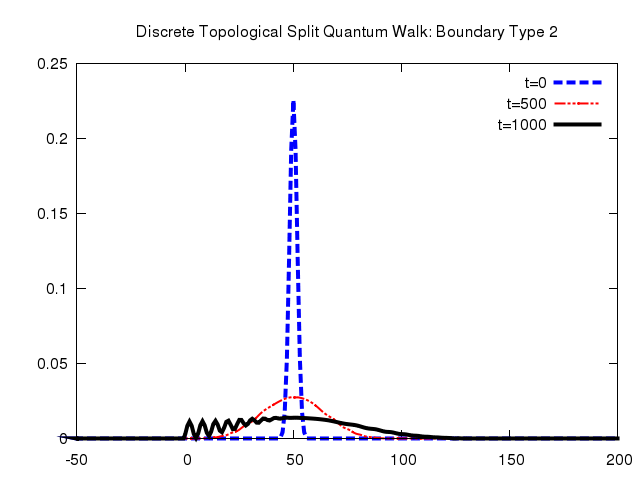}
	\caption{State being reflected at the boundary of topological phases \emph{I} and \emph{III} of a continuous simple-step walk.}\label{fig:8}
\end{figure}

Figures \ref{fig:5} and \ref{fig:7} show the behavior of a discrete split-step quantum walk with topological phases \emph{III} for $x\ge 0$ and \emph{IV} for $x<0$. In Fig.~\ref{fig:5}, we see the same two bound states near the boundary of the two phases as in the case of the simple-step walk. This is expected, because the boundary between phases \emph{III} and \emph{IV} is equivalent to the boundary between two topologically distinct phases of a simple-step quantum walk. This  behavior has been demonstrated experimentally by Kitagawa, \emph{et al.}~\cite{Kitagawa2012}. Figure \ref{fig:7} shows the behavior of the split-step quantum walk away from the boundary. In particular, a system whose initial state is entirely within a single topological phase will never leave the region of that phase. When such a state comes in contact with the boundary between two phases, it is reflected back.

Figures \ref{fig:6} and \ref{fig:8} show the behavior of a discrete split-step quantum walk with topological phases \emph{I} for $x\ge 0$ and \emph{III} for $x<0$. Unlike in the previous case, there is a single bound state at the boundary between phases \emph{I} and \emph{III}, at $x=-1$. This has also been observed experimentally~\cite{Kitagawa2012}. Away from the boundary, we obtain the same qualitative behavior as in the previous case, including reflection of the state at the boundary.

\subsection{Continuous-time limit}

The continuous-time limit is obtained as $\theta_1 \to\pm \frac{\pi}{2}$, $\theta_2\to 0$, or $\theta_2 \to\pm \frac{\pi}{2}$, $\theta_1\to 0$. They correspond to the four distinct topological phases listed above, respectively,
\begin{align}
	I (\theta_1,\theta_2) &= (0, \frac{\pi}{2})\\
	II (\theta_1,\theta_2) &= (0, -\frac{\pi}{2})\\
	III (\theta_1,\theta_2) &= (\frac{\pi}{2},0)\\
	IV (\theta_1,\theta_2) &= (-\frac{\pi}{2},0)
\end{align}
\begin{figure}[htp]
	\centering
	\includegraphics[scale=1.5]{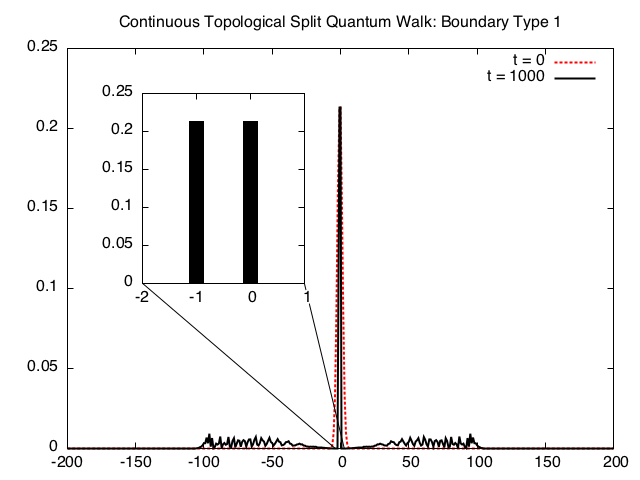}
	\caption{Trapped state at the boundary of topological phases \emph{I} and \emph{II} of a continuous simple-step walk.}\label{fig:9}
\end{figure}

\begin{figure}[htp]
	\centering
	\includegraphics[scale=0.35]{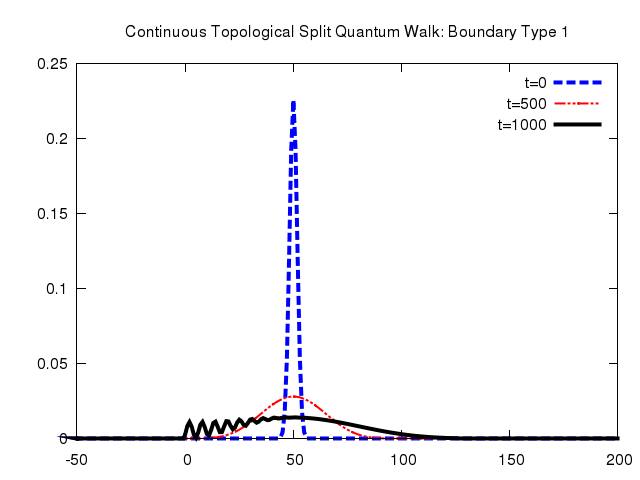}
	\caption{State being reflected at the boundary of topological phases \emph{I} and \emph{II} of a continuous simple-step walk.}\label{fig:10}
\end{figure}
In phase $I$, we set $\theta_1 = \epsilon_1$, $\theta_2 = \frac{\pi}{2} - \epsilon_2$, and consider the scaling limit in which $\epsilon_{1,2}\to 0$, $n\to\infty$, so that the products $n\epsilon_{1,2}$ remain finite. In this limit, $\omega_\pm \to \pm \frac{\pi}{2}$. We have $W^2 = -\mathbb{I} + \mathcal{O} (\epsilon_{1,2})$. We will therefore multiply the wavefunction by a phase $i$, and consider the limit of an even number of steps. Setting
\be \epsilon_{1,2} = \gamma_{1,2} \Delta t \ , \ \ t = 2n\Delta t \ee
we obtain from \eqref{eq8split} the equations of motion in the limit $\Delta t \to 0$,
\bea\label{eqomI} \frac{\partial\Psi_0 (x)}{\partial t} &=& -2\gamma_1 \Psi_1 (x) - \gamma_2 \left[ \Psi_1 (x-1)   + \Psi_1 (x+1) \right] \nonumber\\
\frac{\partial\Psi_1(x)}{\partial t} &=& 2\gamma_1 \Psi_0 (x) + \gamma_2 \left[  \Psi_0 (x-1)    + \Psi_0 (x+1) \right]\ \ \ \
\eea
Working similarly, we obtain the same equations of motion \eqref{eqomI} in the continuous-time limit in phase $II$.

Defining
\be\label{eq47} \Phi_\pm (x) = \pm i \Psi_0 (x) + \Psi_1 (x-1) \ee
we obtain the decoupled equations of motion,
\be\label{eq48} \frac{\partial\Phi_\pm (x)}{\partial t} = \pm 2i\gamma_2 \Phi_\pm (x) \pm i \gamma_1 \left[ \Phi_\pm (x-1)   + \Phi_\pm (x+1) \right] \ee
In phase $III$, we obtain the equations of motion
\bea\label{eqomIII} \frac{\partial\Psi_0 (x)}{\partial t} &=&  \gamma_1 \left[ \Psi_1 (x) +\Psi_1 (x-2) \right] +2\gamma_2 \Psi_1 (x-1)\nonumber\\
\frac{\partial\Psi_1(x)}{\partial t} &=&  - \gamma_1 \left[ \Psi_0 (x) +\Psi_0 (x+2)   \right] - 2\gamma_2 \Psi_0 (x+1) \ \ \ \ \
\eea
and in phase $IV$,
\bea\label{eqomIV} \frac{\partial\Psi_0 (x)}{\partial t} &=&  \gamma_1 \left[ \Psi_1 (x) +\Psi_1 (x+2) \right] +2\gamma_2 \Psi_1 (x+1)\nonumber\\
\frac{\partial\Psi_1(x)}{\partial t} &=&  - \gamma_1 \left[ \Psi_0 (x) +\Psi_0 (x-2)   \right] - 2\gamma_2 \Psi_0 (x-1) \ \ \ \ \
\eea
They can be put into the decoupled form \eqref{eq48}, if we define $\Phi_\pm$ as in \eqref{eq47} in phase $III$, and $\Phi_\pm (x) = \pm i \Psi_0 (x) + \Psi_1 (x+1)$ in phase $IV$.

There are six different boundaries, but only two are qualitatively different. We proceed to consider a representative from each type.

For a system in phase $III$ for $x\ge 0$, and phase $IV$ for $x<0$,
%
working as before, we obtain for $x\le -3$  the equations of motion \eqref{eqomIII}, and for $x\ge 1$, the equations of motion \eqref{eqomIV}. Near the boundary of the two phases, we have
	\bea\label{eq53}
	\frac{\partial\Psi_0 (-2)}{\partial t} &=& \gamma_1 \Psi _1(-2) + 2\gamma_2\Psi _1(-1)\nonumber\\
	\frac{\partial\Psi_1 (-2)}{\partial t} &=& -\gamma_1\left( \Psi _0(-2) + \Psi_0(-4) \right) - 2\gamma_2\Psi _0(-3)\nonumber\\
		\frac{\partial\Psi_0 (-1)}{\partial t} &=& 0\nonumber\\
		\frac{\partial\Psi_1 (-1)}{\partial t} &=& -\gamma_1\Psi_0(-3) - 2\gamma_2\Psi_0(-2)\nonumber\\
		\frac{\partial\Psi_0 (0)}{\partial t} &=& 0 \nonumber\\
		\frac{\partial\Psi_1 (0)}{\partial t} &=& -\gamma_1\Psi _0(2) - 2\gamma_2\Psi_0(1)\nonumber\\
		\frac{\pa\Psi_0(1)}{\pa t} &=& \gamma_1\Psi_1(1) + 2\gamma_2\Psi_1(0) \nonumber\\
		\frac{\pa\Psi_1(1)}{\pa t} &=& -\gamma_1\left( \Psi_0(1) + \Psi_0(3) \right) - 2\gamma_2\Psi_0(2)
	\eea

\begin{figure}[htp]
	\centering
	\includegraphics[scale=1.5]{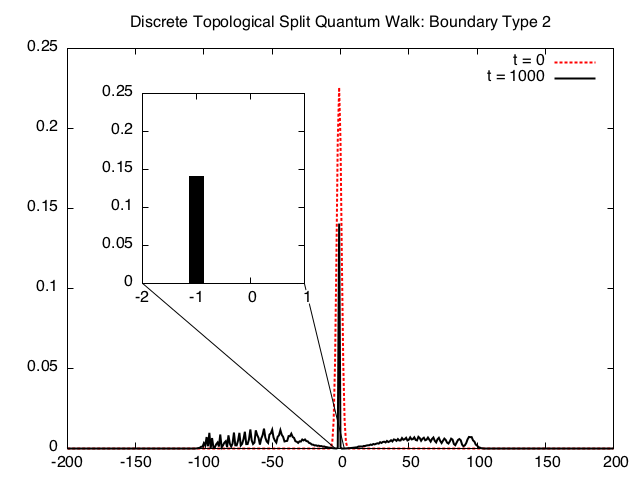}
	\caption{Trapped state at the boundary of topological phases \emph{I} and \emph{III} of a continuous simple-step walk.}\label{fig:11}
\end{figure}

\begin{figure}[htp]
	\centering
	\includegraphics[scale=0.35]{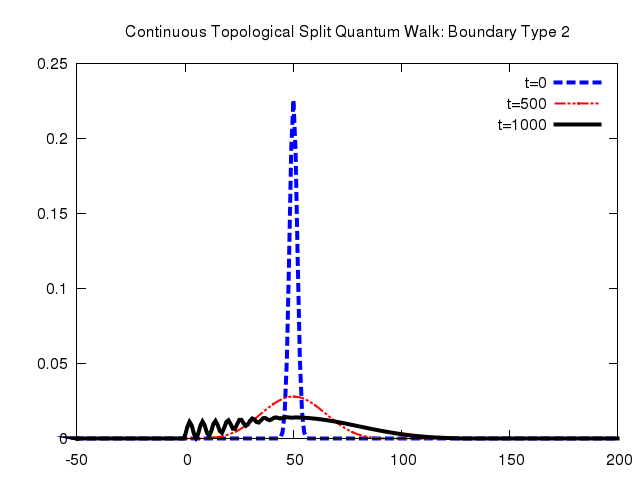}
	\caption{State being reflected at the boundary of topological phases \emph{I} and \emph{III} of a continuous simple-step walk.}\label{fig:12}
\end{figure}

Similarly, for a system in phase $I$ for $x\ge 0$, and phase $III$ for $x<0$,
we obtain for $x\ge 1$, the equations of motion \eqref{eqomI}, and for $x\le -3$, the equations of motion \eqref{eqomIII}.
Near the boundary, we have
\bea\label{eq54}
\frac{\partial\Psi_0 (-2)}{\partial t} &=& \gamma_1\left[\Psi _1(-4)+\Psi _1(-2)\right] + 2\gamma_2\Psi _1(-3) \nonumber\\
\frac{\partial\Psi_1 (-2)}{\partial t} &=& -\gamma_1\Psi _0(-2)-2\gamma_2 \Psi _0(-1)\nonumber\\
\frac{\partial\Psi_0 (-1)}{\partial t} &=& \gamma_1\Psi _1(-3) +2\gamma_2 \Psi _1(-2)\nonumber\\
\frac{\partial\Psi_1 (-1)}{\partial t} &=& 0 \nonumber\\
\frac{\partial\Psi_0 (0)}{\partial t} &=& - \gamma_2\Psi _1(1)\nonumber\\
\frac{\partial\Psi_1 (0)}{\partial t} &=&  \gamma _2\Psi _0(1)\nonumber\\
\eea
Figures \ref{fig:9}, \ref{fig:10}, \ref{fig:11}, and \ref{fig:12} depict the continuous-time limit of the discrete quantum walks shown in Figs. \ref{fig:5}, \ref{fig:6}, \ref{fig:7}, and \ref{fig:8}, respectively. As expected, the observed behavior matches the asymptotic behavior at and away from the boundary in the discrete case. It should be noted that in the continuous-time limit, the bound states can be found analytically from \eqref{eq53} for the boundary between \emph{III} and \emph{IV}, and \ref{eq54} for the boundary between phases \emph{I} and \emph{III}. These bound states are all in agreement with the asymptotic results obtained above in the corresponding discrete cases, as well as experimental results \cite{Kitagawa2012}.

\begin{widetext}

\begin{figure}[htp]
    \centering
	\includegraphics[scale=1.0]{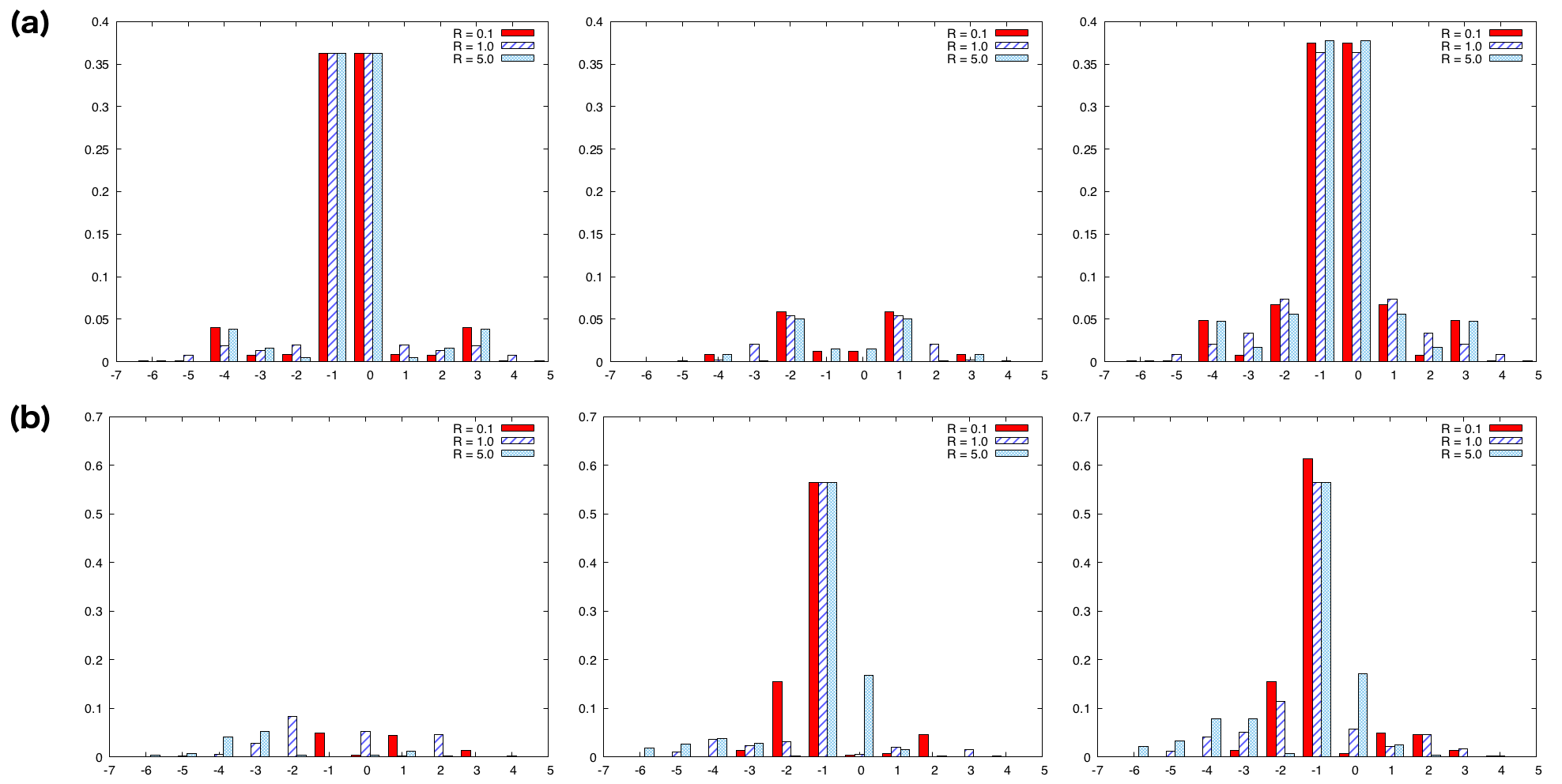}
	\caption{Bound states near the boundary between two phases. From left to right, the distributions of $|\Psi_0(x)|^2, |\Psi_1(x)|^2, |\Psi_0(x)|^2 + |\Psi_1(x)|^2$ after a short time $t=25$. In each plot, the value $R$ corresponds to the ratio between walk parameters $\gamma_1, \gamma_2$. (a) At the boundary between phases \emph{I} and \emph{II} with the initial state centered at $\Psi_0(-1)\ \text{and}\ \Psi_0(0)$. (b) At the boundary between phases \emph{I} and \emph{III} with the initial state centered at $\Psi_1(-1)$}\label{fig:13}
\end{figure}

\end{widetext}

As discussed above, the split-step quantum walk with boundary between the topological phases \emph{III} and \emph{IV} gives rise to two topologically protected bound states at the boundary. In Fig.~\ref{fig:13}, we show that these bound states are robust against small changes in the quantum walk parameters.

\begin{figure}[htp]
    \centering
	\includegraphics[scale=0.75]{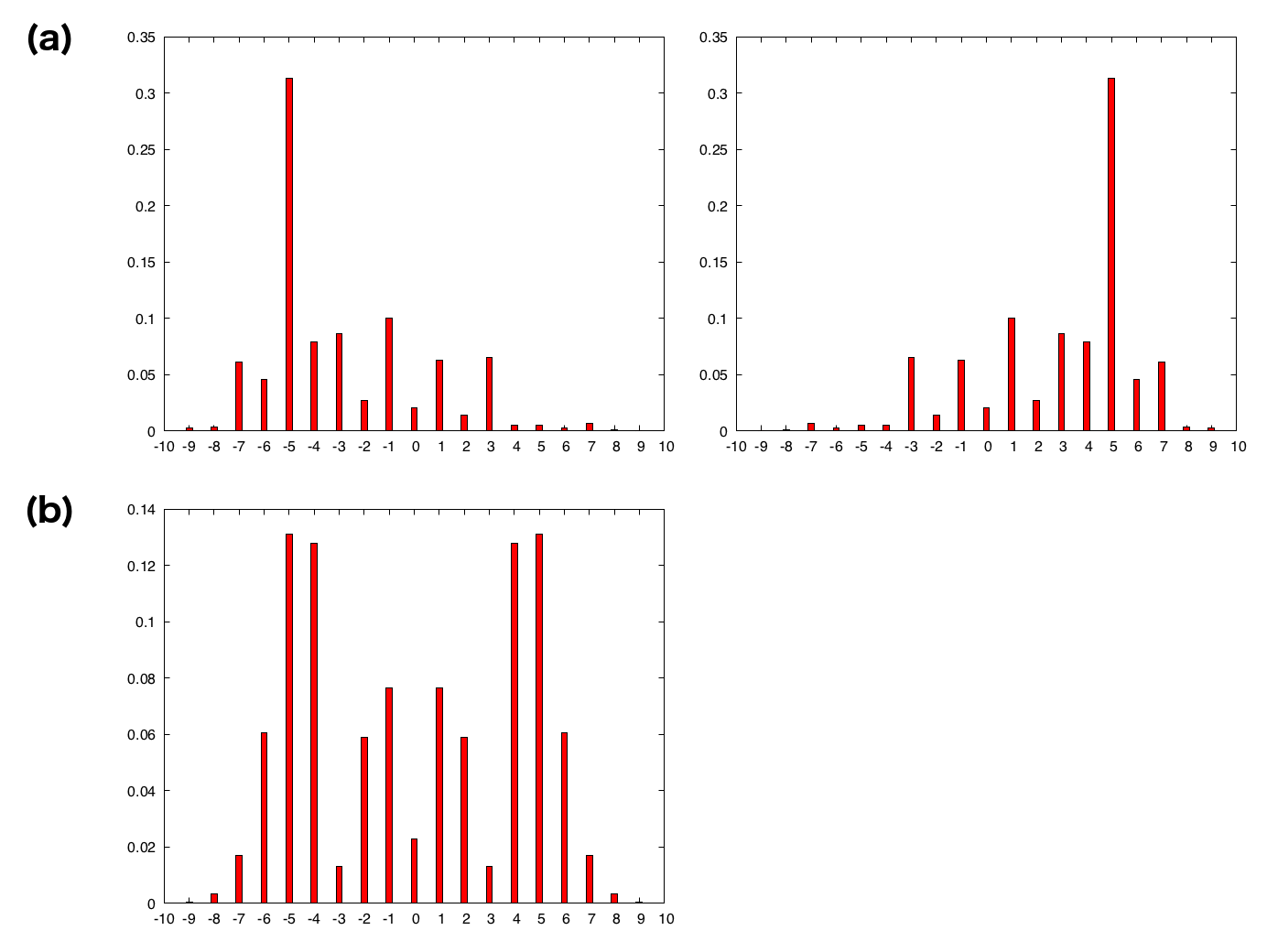}
	\caption{Ballistic diffusion of continuous walks without boundary. (a) In phase \emph{III} with no boundary the quantum walk diffuses ballistically. This diffusion can occur to the right or to the left depending on the initial state. Here the initial state was centered at $\Psi_0(0)$ and $\Psi_1(0)$ with $\Psi_1(0) = \pm \Psi_0(0)$. (b) In phase \emph {I} with no boundary the quantum walk diffuses ballistically with the same behavior regardless of the initial state. }\label{fig:14}
\end{figure}

Figure \ref{fig:14} shows that in the case of a single topological phase, the state ballistically diffuses away from its initial position. In topological phase \emph{III}, one can choose the initial state in such a away that the diffusion only occurs in a single direction. Due to the linearity of the system, it follows that the initial state can be chosen so that the diffusion occurs in both directions. However, in phase \emph{I}, regardless of the choice of initial state, the diffusion invariably occurs in both directions ballistically.

\section{Conclusions}
\label{sec:3}

In conclusion, we have investigated the continuous-time limit of discrete quantum walks with topological phases. In quantum walks, it is common to consider both discrete and continuous-times. In recent years much interest has been devoted to understanding how discrete-time quantum walks can simulator topological insulators. Here we have shown the existence of a continuous-time limit that preserves their topological phases. We considered both simple-step and split-step walks and derived analytically the equations of motion governing their behaviors. Through our analytical solutions we showed the existence of bound states at the boundary of two phases. We also solved the equations of motion numerically in the bulk. In terms of future work it would be interesting to consider the alternative continuous-limit approach given in \cite{Dheeraj2015} to study topological properties of quantum walks.

\acknowledgments{We thank C.\ M. Chandrashekar for illuminating discussions. D.\ C.\ and G.\ S.\ thank the Army Research Laboratory, where most of this work was performed, for its hospitality and financial support.}



\begin{thebibliography}{00}

\bibitem{Kempe2003} J.~Kempe, \emph{``Quantum random walks: An introductory overview,} Contem.\ Phys.\ \textbf{44}, 307 (2003).

\bibitem{Kendon2006} V.~Kendon, \emph{``Decoherence in quantum walks -- a review,"} Math.\ Struct.\ in Comp.\ Sci.\ \textbf{17}, 1169 (2006).

\bibitem{SV} S.\ E.\ Venegas-Andraca, \emph{``Quantum walks: a comprehensive review",} Quant.\ Inf.\ Process.\ \textbf{11}, 1015 (2012).

\bibitem{Childs2010} A.~M.~Childs, \emph{``On the Relationship Between Continuous- and Discrete-Time Quantum Walk,"} Commun.\ Math.\ Phys.\ \textbf{294}, 581 (2010).

\bibitem{Mohseni2008} M.~Mohseni, P.~Rebentrost, S.~Lloyd and A.~Aspuru-Guzik, \emph{``Environment-assisted quantum walks in photosynthetic energy transfer,"} J.\ Chem.\ Phys.\ \textbf{129}, 174106 (2008).

\bibitem{HEN} I.\ Hen and A.\ P.\ Young, \emph{``Exponential complexity of the quantum adiabatic algorithm for certain satisfiability problems,"} Phys.\ Rev.\ E \textbf{84}, 061152 (2011).

\bibitem{Child}A.\ M.\ Childs, D.\ Gosset, and Z.\ Webb, \emph{``Universal Computation by Multiparticle Quantum Walk,"} Science \textbf{339}, 791 (2013).

\bibitem{Schreiber2010} A.~Schreiber, K.~N.~Cassemiro, V.~Potocek, A.~Gabris, P.~J.~Mosley, E.~Andersson, I.~Jex, and C.~Silberhorn, \emph{``Photons Walking the Line: A Quantum Walk with Adjustable Coin Operations,"} Phys.\ Rev.\ Lett.\ \textbf{104}, 050502 (2010).

\bibitem{Broome2010} M.~A.~Broome, A.~Fedrizzi, B.~P.~Lanyon, I.~Kassal, A.~Aspuru-Guzik, and A.~G.~White, \emph{``Discrete Single-Photon Quantum Walks with Tunable Decoherence,"} Phys.\ Rev.\ Lett.\ \textbf{104}, 153602 (2010).

\bibitem{Peruzzo2010} A.~Peruzzo, \textit{et al.}, \emph{``Quantum Walks of Correlated Photons,"} Science \textbf{329}, 1500 (2010).

\bibitem{Schreiber2012} A.~Schreiber, A.~G\'abris, P.~P.~Rohde, K.~Laiho,
M.~\v{S}tefa\v{n}\'ak, V.~Poto\v{c}ek, C.~Hamilton, I.~Jex, and C.~Silberhorn, \emph{``A 2D Quantum Walk Simulation of Two-Particle Dynamics,"} Science \textbf{336}, 55 (2012).

\bibitem{Sansoni2012} L.~Sansoni, F.~Sciarrino, G.~Vallone, P.~Mataloni,
A.~Crespi, R.~Ramponi, and R.~Osellame, \emph{``Two-Particle Bosonic-Fermionic Quantum Walk via Integrated Photonics,"} Phys.\ Rev.\ Lett.\
\textbf{108}, 010502 (2012).

\bibitem{Jeong2013} Y.-C.~Jeong, C.~D.~Franco, H.-T.~Lim, M.~Kim, and Y.-H.~Kim, \emph{``Experimental realization of a delayed-choice quantum walk,"} Nat.\ Comm.\ \textbf{4}, 2471 (2013).

\bibitem{Zahringer2010} F.~Zahringer, G.~Kirchmair, R.~Gerritsma, E.~Solano,
R.~Blatt, and C.~F.~Roos, \emph{``Realization of a Quantum Walk with One and Two Trapped Ions,"} Phys.\ Rev.\ Lett.\ \textbf{104},
100503 (2010).

\bibitem{Schmitz2009} H.~Schmitz, R.~Matjeschk, C.~Schneider, J.~Glueckert,
M.~Enderlein, T.~Huber, and T.~Schaetz, \emph{``Quantum Walk of a Trapped Ion in Phase Space,"} Phys.\ Rev.\ Lett.\
\textbf{103}, 090504 (2009).

\bibitem{Karski2009} M.~Karski, L.~F\"orster, J.-M.~Choi, A.~Steen, W.~Alt,
D.~Meschede, and A.~Widera, \emph{``Quantum Walk in Position Space with Single Optically Trapped Atoms,"} Science \textbf{325}, 174 (2009).

\bibitem{Genske2013} M.~Genske, W.~Alt, A.~Steen, A.~H.~Werner, R.~F.~Werner, D.~Meschede, and A.~Alberti, \emph{``Electric Quantum Walks with Individual Atoms,"} Phys.\ Rev.\ Lett.\
\textbf{110}, 190601 (2013).

\bibitem{Hasan2010} M.~Z.~Hasan and C.~L.~Kane, \emph{``Colloquium: Topological insulators,"} Rev.\ Mod.\ Phys.\ \textbf{82}, 3045 (2010).

\bibitem{Qi2011} X.~Qi and S.~Zhang, \emph{``Topological insulators and superconductors,"} Rev.\ Mod.\ Phys.\ \textbf{83}, 1057 (2011).

\bibitem{Kitagawa2012} T.~Kitagawa, M.~A.~Broome, A.~Fedrizzi, M.~S.~Rudner, E.~Berg,
I.~Kassal, A.~Aspuru-Guzik, E.~Demler, and A.~G.~White, \emph{``Observation of topologically protected bound states in photonic quantum walks,"} Nat.\ Comm.\ \textbf{3}, 882 (2012).

\bibitem{Kit} T.~Kitagawa, M.~S.~Rudner, E.~Berg, and E.~Demler, \emph{``Exploring topological phases with quantum walks,"} Phys.\ Rev.\ A \textbf{82}, 033429 (2010).

\bibitem{Kitagawa2012b} T.~Kitagawa, \emph{``Topological phenomena in quantum walks: elementary introduction to the physics of topological phases,"} Quantum Inform.\ Process.\ \textbf{11}, 1107 (2012).

\bibitem{Lindner2011} N.~H.~Lindner, G.~Refael, and V.~Galitski, \emph{``Floquet topological insulator in semiconductor quantum wells,"} Nat.\ Phys.\ \textbf{7}, 490 (2011).

\bibitem{Obuse2011} H.~Obuse and N.~Kawakami, \emph{``Topological phases and delocalization of quantum walks in random environments,"} Phys.\ Rev.\ B \textbf{84}, 195139 (2011).

\bibitem{Asboth2012} J.~K.~Asb\'oth, \emph{``Symmetries, topological phases, and bound states in the one-dimensional quantum walk,"} Phys.\ Rev.\ B \textbf{86}, 195414 (2012).

\bibitem{Asboth2013} J.~K.~Asb\'oth and H.~Obuse, \emph{``Bulk-boundary correspondence for chiral symmetric quantum walks,"} Phys.\ Rev.\ B \textbf{88}, 121406(R) (2013).

\bibitem{Ramasesh2016} V.~V.~Ramasesh, E.~Flurin, M.~Rudner, I.~Siddiqi, and N.~Y.~Yao, \emph{``Direct Probe of Topological Invariants Using Bloch Oscillating Quantum Walks,"} arXiv:1609.09504 [quant-ph] (2016).

\bibitem{Rakovszky2015} T.~Rakovszky and J.~K.~Asb\'oth, \emph{``Localization, delocalization, and topological phase transitions in the one-dimensional split-step quantum walk,"} Phys.\ Rev.\ A \textbf{92}, 052311 (2015).


\bibitem{Cardano2016} F.\ Cardano, M.\ Maffei, F.\ Massa, B.\ Piccirillo, C.\ de Lisio, G.\ De Filippis, V.\ Cataudella, E.\ Santamato, and L.\ Marrucci, \emph{``Statistical moments of quantum-walk dynamics reveal topological quantum transitions,"} Nat.\ Comm.\ \textbf{7}, 11439 (2016).

\bibitem{Strauch} F.~W.~Strauch, \emph{``Relativistic quantum walks,"} Phys.\ Rev.\ A \textbf{73}, 054302 (2006).

\bibitem{Kitagawa2010} T.~Kitagawa, E.~Berg, M.~Rudner, and E.~Demler, \emph{``Topological characterization of periodically driven quantum systems,"} Phys.\ Rev.\ B \textbf{82}, 235114 (2010).

\bibitem{Obuse2015} H.~Obuse, J.~K.~Asb\'oth, Y.~Nishimura, and N.~Kawakami, \emph{``Unveiling hidden topological phases of a one-dimensional Hadamard quantum walk,"} Phys.\ Rev.\ B \textbf{92}, 045424 (2015).


\bibitem{Dheeraj2015} M.\ N.\ Dheeraj and T.~A.~Brun, \emph{``Continuous limit of discrete quantum walks,"} Phys.\ Rev.\ A \textbf{91}, 062304 (2015).

\end{thebibliography}
\end{document}